\documentclass[11pt,a4paper]{article}

\usepackage[T1]{fontenc}
\usepackage[utf8]{inputenc}
\usepackage{lmodern}
\usepackage{microtype}
\usepackage{setspace}
\usepackage{geometry}
\usepackage{amsmath,amssymb,amsthm}
\usepackage{booktabs}
\usepackage{array}
\usepackage{longtable}
\usepackage{multirow}
\usepackage{graphicx}
\usepackage{caption}
\usepackage{subcaption}
\usepackage{float}
\usepackage{enumitem}
\usepackage{parskip}
\PassOptionsToPackage{hyphens}{url}
\usepackage{hyperref}
\usepackage{natbib}
\usepackage{xcolor}
\usepackage{mdframed}
\usepackage{tikz}
\usepackage{pgfplots}
\pgfplotsset{compat=1.18}
\usetikzlibrary{arrows.meta, positioning, shapes.geometric, fit, backgrounds}

\theoremstyle{definition}
\newtheorem{definition}{Definition}
\newtheorem{proposition}{Proposition}
\newtheorem{remark}{Remark}

\newcommand{\SCI}{\mathrm{SCI}}
\newcommand{\VR}{\mathrm{VR}}
\newcommand{\TS}{\mathrm{TS}}
\newcommand{\HHI}{\mathrm{HHI}}

\hypersetup{
  colorlinks = true,
  linkcolor  = black,
  citecolor  = black,
  urlcolor   = black,
  pdfauthor  = {},
  pdftitle   = {Price as Focal Point}
}
\begin{document}

\hypersetup{pageanchor=false}
\begin{titlepage}
  \centering
  \vspace*{1cm}

  {\LARGE\setstretch{1} Price as Focal Point: Prediction Markets,\\
  Conditional Reflexivity, and the\\
  Politics of Common Knowledge\par}

  \vspace{1cm}

  {\large Maksym Nechepurenko\textsuperscript{*}\par}


  {\normalsize \today\par}

  \vspace{1cm}

  \begin{abstract}
    \noindent\setstretch{1}
    Prediction markets are widely treated as forecasting devices --- instruments that reveal collective expectations about uncertain futures.
    This article argues that under specifiable conditions they also function as \emph{coordination mechanisms}: public probabilities that organize the behavior of voters, donors, journalists, traders, and institutions in ways that can be self-fulfilling or self-defeating.
    Most existing work asks whether prediction markets forecast accurately; this
    paper asks whether accurate forecasting is even the right criterion for
    evaluating a market that has become a public coordination device.
    The core analytical contribution is identifying when this transformation occurs.
    Drawing on recent transaction-level evidence from the 2024 U.S.\ presidential election market, we show that the social force of a market signal depends less on its size than on its \emph{persistence}, the breadth of responding trader types, and the degree of cross-platform consensus.
    We introduce a Signal Credibility Index (SCI) --- combining the variance ratio
    $\VR(6)$, a two-sidedness diagnostic, and a trader-concentration adjustment ---
    as a microstructure-grounded criterion for predicting when price moves will acquire behavioral traction.
    Applying this framework to three major 2024 political shocks, we demonstrate that superficially similar events generated qualitatively distinct signal types with different implications for elite coordination.
    A cross-platform comparison reveals that the most socially visible market produced the least accurate forecasts, establishing a systematic decoupling of social authority from epistemic robustness.
    The framework has direct implications for the regulation of prediction markets as public information infrastructure: the deregulatory trajectory of 2025--2026 may improve liquidity while systematically degrading the epistemic quality of the public signals that now organize elite behavior.
  \end{abstract}

  \vspace{0.5cm}

  \begin{flushleft}
    \setstretch{1.0}
    perceived inevitability; information aggregation.

    \vspace{0.25cm}

  \end{flushleft}

  \vfill
  \begin{flushleft}
    {\footnotesize\textsuperscript{*} Research Department, Devnull FZCO, Dubai, UAE; E-mail: \href{mailto:maksym@devnull.ae}{maksym@devnull.ae}.\par}
  \end{flushleft}
\end{titlepage}
\hypersetup{pageanchor=true}

\newpage
\pagenumbering{arabic}

\section*{Introduction}
\addcontentsline{toc}{section}{Introduction}

Prediction markets were once easy to describe: something happens in the world,
traders absorb the news, and prices move to reflect revised expectations.
In that familiar model, markets sit downstream of reality.
They look like mirrors of events, not participants in them
\citep{snowberg2013prediction}.

That description is now incomplete.
A platform such as Polymarket does not merely aggregate beliefs about elections,
geopolitical shocks, legal decisions, or macroeconomic events.
By publishing continuously updated probabilities in a public, legible, and
financially coded form, it can also shape how those events are interpreted by
voters, donors, investors, journalists, and institutions.
The relationship is no longer simply events-to-markets.
It is increasingly recursive: markets influence beliefs, beliefs influence
behavior, and behavior feeds back into outcomes
\citep{ng2026price,reichenbach2025polymarket}.

That possibility is worth taking seriously precisely because prediction markets
have earned genuine forecasting credibility.
In a long-run study of the Iowa Electronic Markets, \citet{berg2008prediction}
found that market forecasts outperformed polls in 451 of 596 comparisons beyond
the 100-day horizon.
But that credibility attaches to prediction markets \emph{as a class}, not
automatically to every platform that now trades under the label.
Recent comparative evidence challenges the assumption:
across the 2024 election cycle, the platform with the highest visibility and
the deepest trading pools was the \emph{least} accurate by standard forecasting
metrics \citep{clinton2025accuracy}.
Social authority and epistemic robustness are systematically decoupled --- and
that decoupling is this paper's central observation.

The deeper claim is structural, not merely empirical.
\emph{Most existing work asks whether prediction markets forecast accurately;
this paper asks whether accurate forecasting is even the right criterion for
evaluating a market that has become a public coordination device.}
Once a visible price is interpreted as the best available summary of informed
expectation, it begins to alter the environment it is supposedly measuring.
The key shift is from dispersed private belief to public common knowledge.
When a market expresses scattered private judgments in a single salient number,
those judgments become a focal signal around which behavior can coordinate
\citep{rothschild2014polls}.

We proceed in four analytical moves.
First, we distinguish one-directional from reflexive market systems, and
formalize the coordination condition that marks the transition between them
(Sections~1--2).
Second, we examine the microstructure conditions under which a price move
acquires social force, introducing the Signal Credibility Index and documenting
an authority paradox in the 2024 data (Sections~3--4).

\begin{figure}[H]
  \centering
  \includegraphics[width=0.82\textwidth]{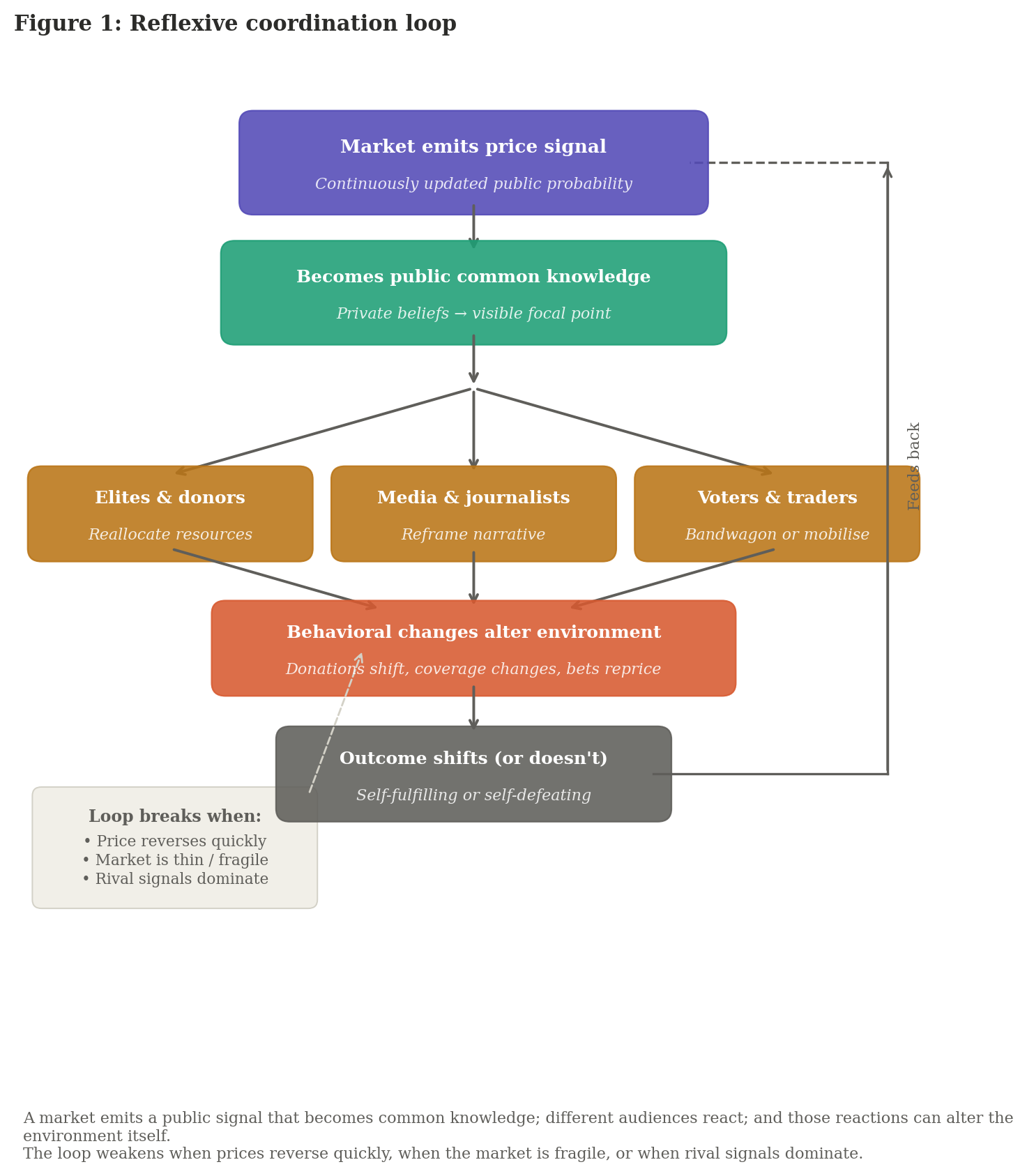}
  \caption{Reflexive coordination loop. A market emits a public signal,
    that signal becomes common knowledge, different audiences react, and
    those reactions can alter the environment itself. The loop weakens
    when prices reverse quickly, when the market is fragile, or when
    rival signals dominate.}
  \label{fig:loop}
\end{figure}

Third, we trace the behavioral consequences of credible public signals, covering
self-fulfilling loops, self-defeating dynamics, and the particular potency of
perceived inevitability (Sections~5--7).
Fourth, we draw out implications for the governance of prediction markets as
democratic infrastructure (Sections~8--11).

\section{From One-Directional to Reflexive Systems}
\label{sec:reflexive}

A one-directional model treats the market as a passive processor of external
facts.
A debate performance improves a candidate's prospects; the contract reprices
upward.
A ceasefire becomes less likely; the relevant market falls.
Information moves from the world into the market, and the market's role is to
summarize rather than intervene \citep{snowberg2013prediction}.

That model is often useful, but in social systems it is incomplete.
Political actors, firms, media organizations, and bureaucracies do not merely
observe public probabilities; they respond to them.
Once that happens, the market is no longer just a scoreboard attached to events
from the outside.
It becomes part of the informational environment within which subsequent
decisions are made \citep{rothschild2014polls,manski2006interpreting}.

This is the point at which reflexivity enters.
In \citeauthor{soros1987alchemy}'s formulation \citep{soros1987alchemy}, beliefs are not only
representations of the world; they can also affect the world they represent.
Prediction markets are especially potent carriers of this dynamic because they
compress diffuse and heterogeneous expectations into a single public number that
appears disciplined by money.
A claim such as ``Candidate~$X$ has an 80\% chance'' is not merely descriptive.
It is a highly usable social signal: it tells observers not just what may happen,
but what other observers with capital appear to think will happen
\citep{manski2006interpreting,wolfers2004prediction}.

The mechanism is therefore not best understood as truth versus error, but as
private belief versus public common knowledge.
A population can contain many private expectations without those expectations
having large behavioral effects, because no one knows whether others share them
strongly enough to act.
A market price changes that.
It can transform scattered judgment into a focal signal that actors can
coordinate around, even if the underlying information set is imperfect
\citep{rothschild2014polls}.

The macroeconomic case is instructive and is developed fully in Section 5; the short version is that after a hotter-than-expected inflation print in April 2024, CME FedWatch's implied probability of a June Federal Reserve rate cut fell from roughly 50\% to 19.4\% in a single day \citep{reuters2024fedwatch} — and that shift became a public reference point for pricing, commentary, and positioning regardless of its epistemic precision.

The key insight from \citet{chernov2025comovement} sharpens this picture.
Their structural model of the 2024 presidential election identifies a
\emph{two-factor} structure driving the vast majority of variation in voter
preferences across states: one factor captures macroeconomic fundamentals, and
a second captures event-driven sentiment shifts.
Prediction markets that reprice along the second factor --- as they clearly did
after the June debate and the July assassination attempt --- are not pricing
one isolated probability.
They are reweighting a nationally correlated belief structure that loads
simultaneously on many states.
This amplifies the coordination power of a price move: a single visible number
shifts not one voter's assessment but a correlated cluster of assessments across
the electoral map.

None of this establishes a universal self-fulfilling mechanism.
\citet{rothschild2014polls} show that the behavioral effects of public forecasts
are mixed, context-dependent, and often overstated.
The relevant claim is therefore explicitly conditional.
Those scope conditions can be stated directly: reflexivity is strongest when a
market signal is public and widely legible, when the price move is persistent
rather than fleeting, when relevant actors are coordinating under uncertainty,
and when outsiders treat the market as credible.
It is weaker when price moves reverse quickly, when alternative signals dominate
the interpretive environment, or when audiences do not treat the market as
especially authoritative.
The question is not whether prediction markets always move behavior, but under
what conditions they add enough social force to matter.

\section{A Formal Framework for Conditional Reflexivity}
\label{sec:formal}

\subsection{The Basic Coordination Condition}
\label{subsec:coordination-condition}

Let $p_t \in [0,1]$ denote the market-implied probability at time $t$, and let
$O_t \in \mathbb{R}$ denote the underlying outcome variable --- for example, a
candidate's actual electoral support, a central bank's policy decision, or a
geopolitical outcome.
In a purely one-directional system, the outcome is a function only of
fundamental drivers $\mathbf{X}_t$ and a noise term $\varepsilon_t$:
\begin{equation}
  O_t = f(\mathbf{X}_t,\, \varepsilon_t), \qquad \frac{\partial O_t}{\partial p_t} = 0.
  \label{eq:one-directional}
\end{equation}

In a reflexive system, the market price enters directly into the causal chain
because actors observe $p_t$ and update their behavior accordingly:
\begin{equation}
  O_t = f\!\left(\mathbf{X}_t,\, p_t,\, \varepsilon_t\right), \qquad
  \frac{\partial O_t}{\partial p_t} \neq 0.
  \label{eq:reflexive}
\end{equation}

\begin{definition}[Coordination condition]
  A market probability $p_t$ is \emph{socially consequential} if and only if
  the magnitude of the reflexive channel exceeds a minimum behavioral threshold
  $\delta > 0$:
  \begin{equation}
    \left|\frac{\partial O_t}{\partial p_t}\right| \cdot \sigma(p_t) > \delta,
    \label{eq:coord-condition}
  \end{equation}
  where $\sigma(p_t)$ is the volatility of the price signal over the relevant
  horizon.
\end{definition}

This formalization makes the empirical question precise: the coordination thesis
requires estimating $\partial O_t/\partial p_t$, not merely showing that $p_t$
moves.
The baseline case ($\partial O_t/\partial p_t = 0$) is the null hypothesis, and
the threshold $\delta$ separates noise from meaningful behavioral influence.

\subsection{Audience Heterogeneity and the Coordination Channel}
\label{subsec:audience}

The aggregate reflexive effect is the sum of responses across heterogeneous
audiences.
Let $i \in \{\text{elites},\, \text{media},\, \text{voters},\, \text{traders}\}$
index audience types.
Each audience has a response function $\beta_i(p_t, \sigma_i)$ that depends on
the observed probability and that audience's sensitivity $\sigma_i$ to market
signals.
The total behavioral response is:
\begin{equation}
  B_t = \sum_{i} \omega_i \cdot \beta_i(p_t,\, \sigma_i),
  \label{eq:aggregate-response}
\end{equation}
where $\omega_i$ is the weight of audience $i$ in the outcome-generating
process.
The literature is clear on the ordering: $\sigma_{\text{elites}} \gg
\sigma_{\text{voters}}$ \citep{rothschild2014polls,farjam2021bandwagon}.
Elite and media responses are the primary channel through which market
probabilities become socially consequential; direct mass-voter effects are the
exception rather than the rule.

\subsection{The Market-Update Dynamics}
\label{subsec:market-update}

The market itself updates in response to new information and, in a reflexive
system, in response to the behavioral changes $B_t$ has induced:
\begin{equation}
  p_{t+1} = g\!\left(p_t,\, \Delta I_t,\, B_t,\, \nu_t\right),
  \label{eq:market-update}
\end{equation}
where $\Delta I_t$ is new exogenous information, $B_t$ is the aggregate
behavioral response that feeds back into fundamentals, and $\nu_t$ is a
market-microstructure noise term.
The self-fulfilling loop closes when $B_t$ affects $O_t$ in a direction that
validates the original $p_t$; the self-defeating case arises when $B_t$ has
the opposite sign.

\begin{remark}
  Equation~\eqref{eq:market-update} highlights why the microstructure noise
  term $\nu_t$ is non-trivial.
  In a thin market, $\nu_t$ can be large enough to produce large $p_{t+1}$
  moves that are informationally vacuous but socially consequential if treated
  as credible.
  Section~\ref{sec:microstructure} develops this point formally.
\end{remark}

\section{Why Probabilities Behave Like Social Signals}
\label{sec:social-signals}

Probabilities on a prediction market do more than summarize opinion.
They also make expectations public in a form that is easy to coordinate around.
The social effect of a forecast depends not only on whether it is accurate, but
on whether actors treat it as a credible signal of what other actors are likely
to believe and do.

The first mechanism is the conversion of dispersed private judgments into
visible common knowledge.
Many people may independently suspect that a candidate is weakening, that a
central-bank move is less likely, or that a geopolitical event is becoming more
probable.
But private beliefs do not by themselves coordinate action.
A market price does so because it compresses heterogeneous views into a single
number that appears disciplined by incentives and capital, producing not simply
information but a Schelling focal point \citep{schelling1960strategy}.

The two-factor structure identified by \citet{chernov2025comovement} adds
precision to this claim.
They show that variation in state-level voter preferences in the 2024 cycle is
dominated by two latent factors: a \emph{fundamentals factor} (loading on
economic conditions) and an \emph{event factor} (loading on campaign shocks
such as debates, withdrawals, and external incidents).
A prediction market price move driven by the event factor thus represents a
correlated update across many states simultaneously.
Failing to account for this correlation can bias estimated win probabilities by
more than 10 percentage points.
This means that the social power of a single market move is amplified by the
latent correlation structure of electoral opinion: the market is not shifting
one person's posterior, but a nationally correlated belief cluster.

The focal-point effect works differently across audiences.
For \textbf{voters}, the relevant mechanism is bandwagon or viability inference:
if a candidate appears to be losing badly, some marginal voters may shift toward
the perceived winner or disengage from a race that looks settled
\citep{farjam2021bandwagon}.
For \textbf{elites and donors}, the mechanism is more strategic: a public
probability can influence where money, endorsements, staff time, and
institutional support are allocated.
For \textbf{media organizations and commentators}, the mechanism is narrative: a
market number becomes a shorthand for momentum, decline, inevitability, or
crisis.
These channels should not be collapsed, because the evidence is not equally
strong across them.

The Biden episode after the June 27, 2024 debate illustrates the
elite-coordination channel most clearly.
Polymarket odds that Biden would withdraw reportedly rose from approximately
19\% before the debate to 33\% during it and to roughly 50\% by the following
day, with betting across related contracts approaching \$80 million by early
July \citep{wsj2024biden}.
The important point is not that the market caused elite panic.
It is that the market rapidly became part of the informational environment
through which elites interpreted viability.
Once a highly visible public probability moves that quickly, donors, party
actors, and journalists no longer respond only to the debate itself; they also
respond to the fact that ``the market'' now treats withdrawal as a serious live
outcome.

That said, as we show in Section~\ref{sec:microstructure}(see Figure~\ref{fig:paths}), the debate shock is
better understood as a strong example of rapid social salience than as a stable
coordination anchor.
Transaction-level evidence indicates that the debate-induced Trump YES price
jump peaked at approximately $+0.1113$ relative to the pre-event price but
ended the four-hour post-event window only $+0.0200$ above baseline
\citep{tsang2026political}.
The signal was vivid but not durable, and the distinction matters.

A price is influential not because it is guaranteed to be correct, but because
it is legible, public, and easy to cite.
In that sense, the social power of prediction markets comes less from pure
persuasion than from their ability to organize expectations into a common
reference point --- regardless of the epistemic quality of that reference point.

\section{Market Microstructure and Signal Quality}
\label{sec:microstructure}

All of the preceding would be less troubling if market probabilities were always
robust expressions of collective intelligence.
They are not.
The main problem is not that markets can be wrong; it is that observers routinely
treat a market price as a deep, objective probability even when the underlying
signal is structurally fragile.

This section introduces three diagnostics for signal quality, proposes the
Signal Credibility Index as an integrating measure, documents an authority
paradox in the 2024 data, and examines the fragmentation problem that arises
when multiple platforms publish divergent probabilities simultaneously.

\subsection{Microstructure Diagnostics}
\label{subsec:diagnostics}

Following \citet{tsang2026political} and \citet{tsang2026anatomy}, two compact
diagnostics help distinguish durable repricing from temporary pressure.

\paragraph{Variance ratio.}
The variance ratio compares the variance of 30-minute returns to six times the
variance of 5-minute returns:
\begin{equation}
  \VR(6) = \frac{\mathrm{Var}(r_t^{(30\,\mathrm{min})})}{6\,\mathrm{Var}(r_t^{(5\,\mathrm{min})})}
  \label{eq:vr}
\end{equation}
Under a random walk, $\VR(6) = 1$.
Values above unity indicate short-horizon drift --- the repricing is
accumulating, suggesting informed trading.
Values below unity indicate mean-reversion or reversal, consistent with
temporary order-flow pressure.

\paragraph{Two-sidedness index.}
Let $\mathcal{B}_t$ and $\mathcal{S}_t$ denote the signed buy and sell volumes
in a given window. The two-sidedness index is:
\begin{equation}
  \TS_t = 1 - \frac{\lvert \mathcal{B}_t - \mathcal{S}_t \rvert}{\mathcal{B}_t + \mathcal{S}_t}
  \label{eq:twosided}
\end{equation}
Values near 1 indicate balanced buy-and-sell pressure, consistent with
disagreement among informed traders.
Values near 0 indicate strongly one-sided flow, consistent with directional
consensus (or thin-market manipulation).

\subsection{The Signal Credibility Index}
\label{subsec:sci}

We integrate the two diagnostics with a concentration adjustment into a
composite Signal Credibility Index for shock~$s$:
\begin{equation}
  \SCI_s
  = \underbrace{\VR(6)_s}_{\text{persistence}}
    \times
    \underbrace{(1 - \TS_s)}_{\text{consensus}}
    \times
    \underbrace{(1 - \HHI_s)}_{\text{breadth}},
  \label{eq:sci}
\end{equation}
where $\VR(6)_s > 1$ indicates drift (informed repricing), $(1-\TS_s)$ rewards
directional consensus over two-sided disagreement, and $(1 - \HHI_s)$ penalises
concentrated order flow.

More precisely, since high two-sidedness indicates \emph{disagreement} (which
suppresses coordination), and low two-sidedness indicates consensus (which
supports it), the consensus component enters inversely:
\begin{equation}
  \SCI_s = \VR(6)_s \times (1 - \TS_s) \times (1 - \HHI_s),
  \label{eq:sci-expanded}
\end{equation}
where $\HHI_s$ is the Herfindahl-Hirschman index of the distribution of position
sizes among active traders in the post-shock window.
High concentration ($\HHI_s \to 1$) lowers the SCI because a price move driven
by a narrow slice of capital may reflect individual conviction rather than broad
consensus.
Low concentration ($\HHI_s \to 0$) raises the SCI.

\begin{proposition}[Coordination threshold]
  A market signal $\Delta p_s$ following shock $s$ generates socially
  consequential coordination effects if and only if
  $\SCI_s > \tau$, where $\tau > 0$ is a domain-specific threshold
  determined by audience sensitivity $\sigma_i$ and the persistence horizon
  relevant for each behavioral response.
  \label{prop:threshold}
\end{proposition}

Table~\ref{tab:shocks} applies the SCI framework to the three major 2024 shocks
documented in \citet{tsang2026political}.

\begin{table}[H]
  \centering
  \caption{Microstructure diagnostics and predicted social force for three
           political shocks, Polymarket Trump YES contract, 2024}
  \label{tab:shocks}
  \setstretch{1.2}
  \small
  \begin{tabular}{lcccc>{\raggedright\arraybackslash}p{4.2cm}}
  \toprule
  \textbf{Event} & $|\Delta p|_{\text{imm}}$ & $|\Delta p|_{+4\text{h}}$ &
  $\VR(6)$ & Est.\ $\SCI$ & \textbf{Social force} \\
  \midrule
  Debate (Jun 28)        & $+0.111$ & $+0.020$ & $<1$ & $\approx 0.15$ &
    Weak --- vivid but transient \\
  Assassination (Jul 13) & $+0.109$ & $+0.109$ & $>1$ & $\approx 0.68$ &
    Strong --- persistent signal \\
  Biden dropout (Jul 21) & $-0.039$ & $-0.020$ & $\approx 1$ & $\approx 0.22$ &
    Weak --- disputed anchor \\
  \bottomrule
\end{tabular}
  \par\smallskip
  \raggedright\footnotesize
  \textit{Notes:} Values from \citet{tsang2026political} (5-minute event-time bins).
  SCI computed from Equation~\eqref{eq:sci-expanded}; HHI not directly reported and
  treated as moderate across all events.
  $\TS$: qualitative assessment from reported two-sidedness patterns.
  A high-SCI signal is predicted to generate stronger elite coordination,
  greater media reframing, and more durable donor reallocations.
\end{table}

\begin{figure}[H]
  \centering
  \includegraphics[width=\textwidth]{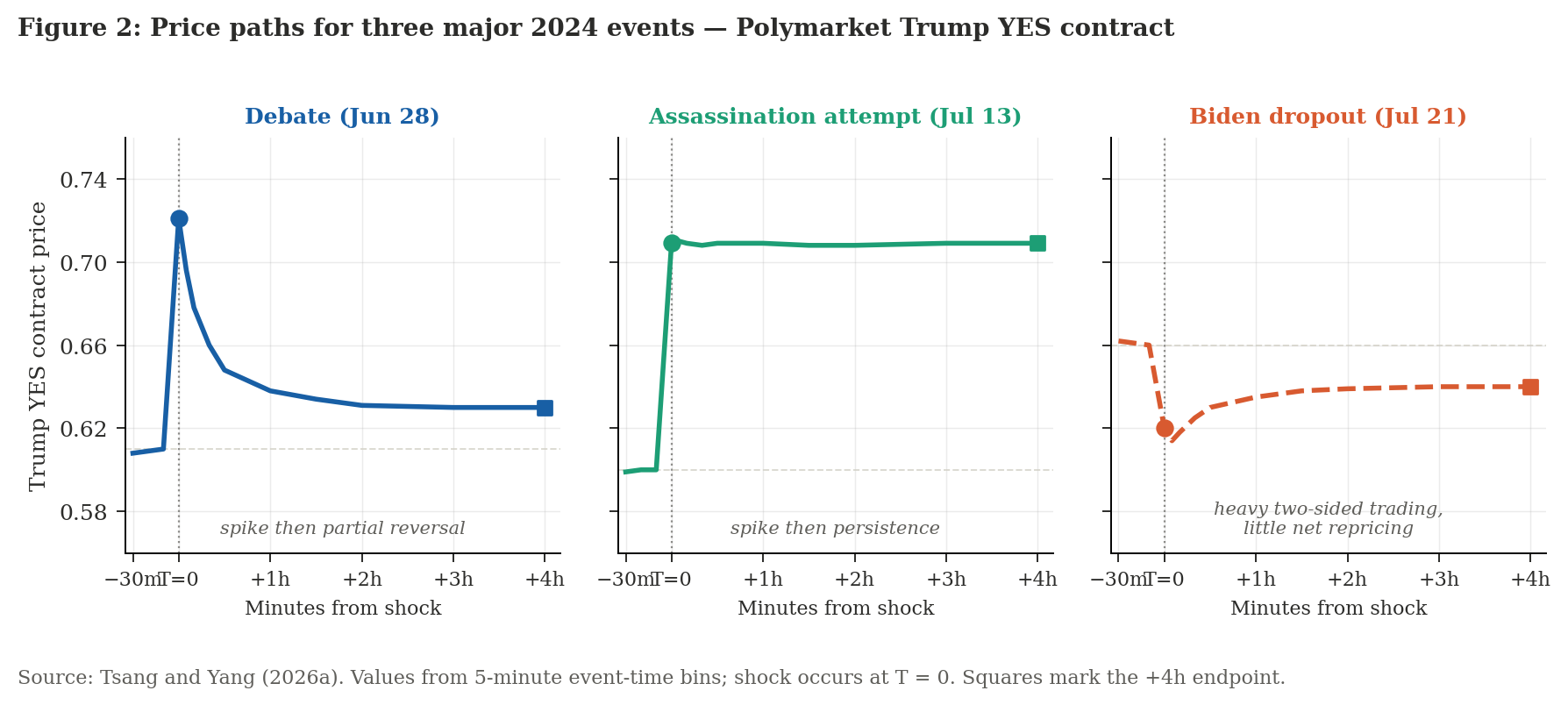}
  \caption{Price paths for three major 2024 events in the Polymarket
    Trump YES contract. The debate produces a spike followed by partial
    reversal; the assassination attempt produces a persistent repricing;
    the Biden dropout produces heavy two-sided trading with little net
    change. Source: \citet{tsang2026political}.}
  \label{fig:paths}
\end{figure}

The debate shock produced a large immediate move ($+0.111$) that largely
reversed within four hours.
Its low SCI predicts weak coordination effects: the signal was socially salient
--- widely cited, rapidly disseminated --- but not epistemically credible to
sophisticated elites who could observe the reversal.
The assassination-attempt shock produced an identically large immediate move
($+0.109$) that \emph{persisted} through the entire post-event window.
Its high SCI predicts strong coordination: this is the case closest to the
article's strongest version of the reflexivity thesis.
The Biden dropout produced a much smaller immediate move with strong two-sided
trading --- indicating genuine disagreement among informed traders about how
to map the event into election odds --- and correspondingly low predicted
social force despite the enormous political salience of the event.

\subsection{The Authority Paradox}
\label{subsec:authority-paradox}

A key implication of the SCI framework is temporal.
\citet{tsang2026anatomy} document that Polymarket's 2024 presidential election
market underwent substantial maturation over its ten-month life.
Early in the cycle, Kyle's $\lambda$ --- the price-impact coefficient, measuring
how much a given order volume moves prices --- was high, indicating thin liquidity.
As the election approached, $\lambda$ declined by more than an order of
magnitude.
Arbitrage deviations narrowed, cross-market participation broadened, and the
market converged toward arbitrage-consistent pricing.

This creates what we call the \textbf{authority paradox}:
\begin{mdframed}
  \textit{A prediction market may acquire social authority --- being widely cited,
  shaping elite interpretations, and organizing donor behavior --- precisely
  when it is least epistemically mature.
  Social credibility is front-loaded; market quality is back-loaded.}
\end{mdframed}

\begin{figure}[H]
  \centering
  \includegraphics[width=0.9\textwidth]{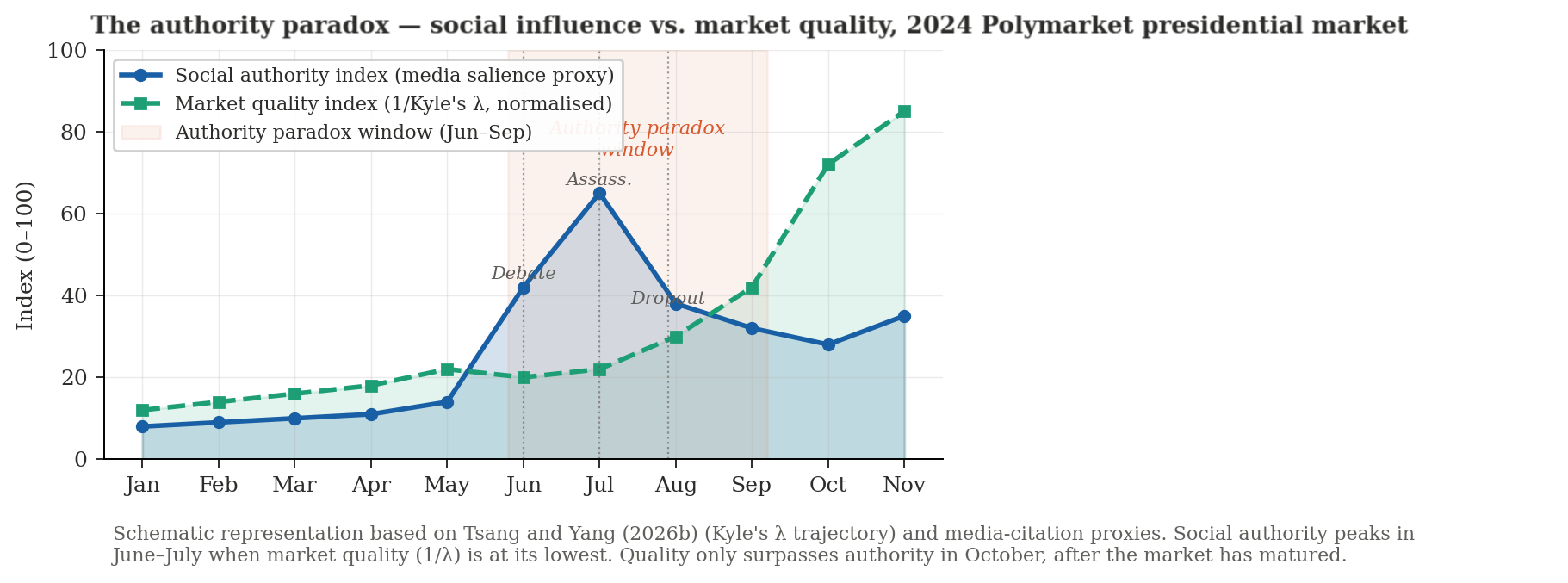}
  \caption{The authority paradox: social influence index versus market
    quality index over the 2024 Polymarket presidential election market.
    Social authority peaks in June--July precisely when Kyle's
    $\lambda$ is highest (market quality lowest). The shaded region
    marks the authority paradox window. Sources: \citet{tsang2026anatomy}
    for the $\lambda$ trajectory; media-citation proxies for the
    authority series.}
  \label{fig:paradox}
\end{figure}

In the 2024 cycle, the market was most influential in June-August, during and
after the Biden-debate crisis.
This is exactly the period when Kyle's $\lambda$ was highest --- when small
order flows could produce large visible moves --- and before the October
maturation brought the market closer to informational efficiency.
A visible probability can acquire public authority early, when epistemic
reliability is still weak, and retain that authority even as the underlying
market improves.
The implication for interpretation is stark: the probability most likely to
influence elite behavior is the one emitted during the period of greatest
structural fragility.

\subsection{Cross-Platform Fragmentation}
\label{subsec:fragmentation}

A deeper complication for the coordination argument is that prediction markets
do not produce a single focal signal.
\citet{ng2026price} provide the first systematic cross-platform comparison of
Polymarket, Kalshi, PredictIt, and Robinhood during the 2024 presidential
election.
Their central finding is that despite covering identical events, these platforms
displayed systematically different probabilities for extended periods.
Specifically, Polymarket led Kalshi in price discovery --- meaning Polymarket
prices predicted future Kalshi prices but not vice versa --- yet meaningful
price disparities persisted even in the most liquid market pair.

This fragmentation matters because the coordination argument depends on the
existence of a \emph{shared} reference point.
If journalists cite Polymarket, traders observe Kalshi, and online political
communities reference PredictIt, then the ``common knowledge'' signal becomes
ambiguous.
Let $p_t^{(k)}$ denote the probability on platform $k$ at time $t$.
The effective common-knowledge signal for coordination purposes is:
\begin{equation}
  p_t^{*} = \sum_k w_k(t) \cdot p_t^{(k)},
  \label{eq:ck-signal}
\end{equation}
where $w_k(t)$ are audience-weighted shares of citation and visibility.
Fragmentation raises the variance of $p_t^{*}$ and lowers its
coordination power: actors who are uncertain which platform's probability is
``the real one'' coordinate less effectively around any single number.

Empirically, \citet{clinton2025accuracy} document that on 62 of 65 days before
the 2024 election, the Harris and Trump contract prices did not sum to \$1.00 on
at least one major platform --- meaning the platforms were internally
inconsistent on a given day, not merely inconsistent with each other.
This is a stringent test of informational quality, and the failure rate is high
enough to matter for any agent relying on the displayed probability as a
decision input.

\begin{figure}[H]
  \centering
  \includegraphics[width=0.85\textwidth]{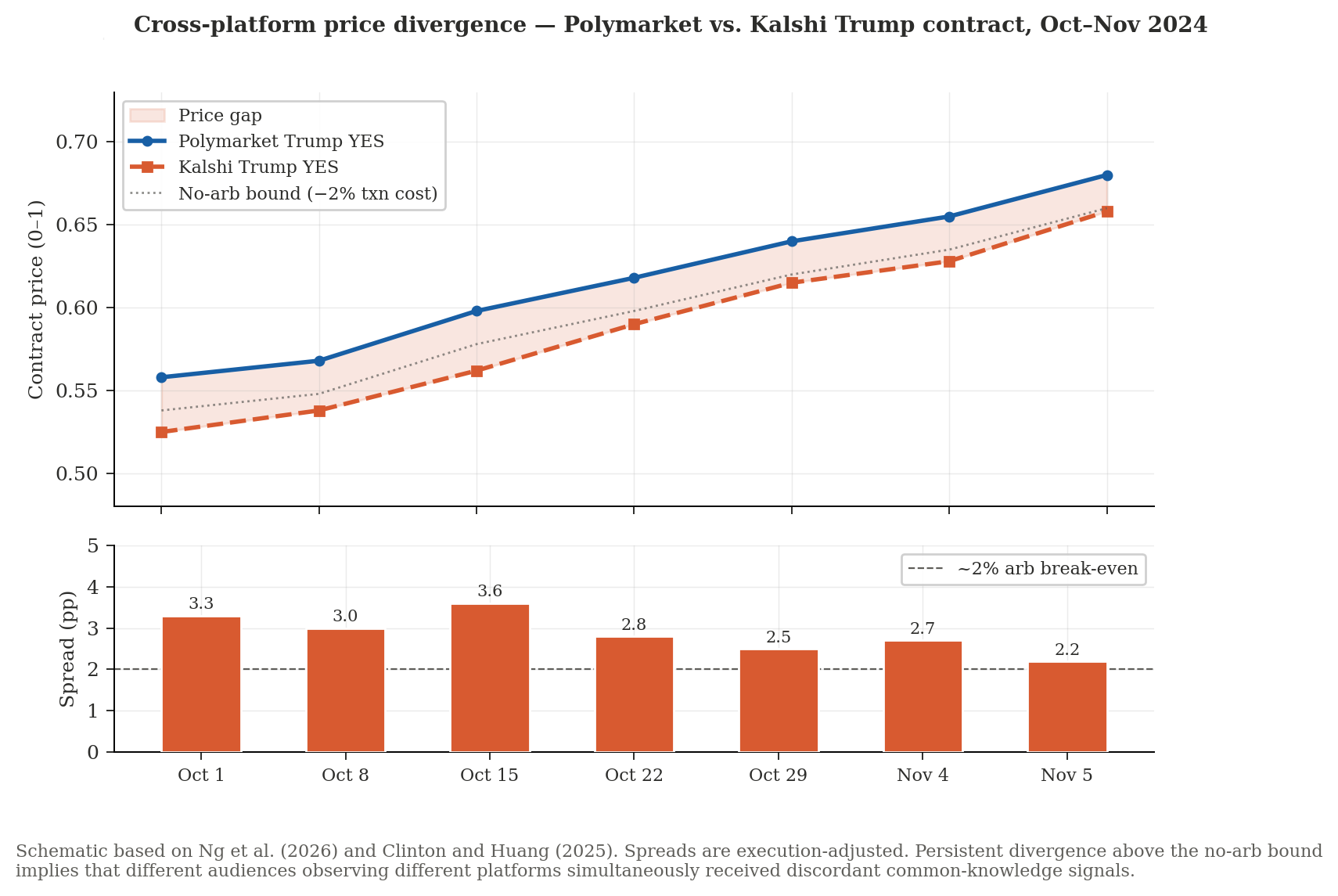}
  \caption{Cross-platform price divergence: Polymarket versus Kalshi
    Trump YES contract, October--November 2024. The lower panel shows
    the weekly spread in percentage points; all observations exceed the
    $\approx$2\% arbitrage break-even, meaning different platforms were
    simultaneously providing discordant common-knowledge signals.
    Sources: \citet{ng2026price}; \citet{clinton2025accuracy}.}
  \label{fig:divergence}
\end{figure}

\subsection{The Composition of Order Flow}
\label{subsec:orderflow}

The informational content of a post-shock price move depends not only on its
size and persistence but on \emph{who} is trading.
\citet{tsang2026political} show that volume rises sharply after major political
shocks, but the composition of that volume is non-uniform.
The response is disproportionately driven by pre-event \emph{incumbent traders}
--- those with substantial existing positions --- who are also more likely to
flip positions after news arrives.
This finding is consistent with inventory management and strategic repositioning,
not just neutral wisdom-of-crowds updating.

\citet{ng2026price} further show that net order imbalance from \emph{large}
trades strongly predicts subsequent returns across platforms.
This dual finding creates an interpretive tension.
Large-trader concentration is simultaneously a source of:
\begin{enumerate}[noitemsep]
  \item \textbf{Fragility}: concentrated capital shapes visible prices in ways
        that may not reflect broad consensus, so outsiders may mistake a
        capital-weighted signal for a democratic one.
  \item \textbf{Informativeness}: large-trader order imbalance predicts future
        returns, suggesting these participants are more informed or faster
        to react than the broader pool.
\end{enumerate}
Observers cannot easily distinguish between informed concentration and
misleading dominance.
This is why the SCI's concentration adjustment ($1 - \HHI_s$) is a
conservative correction: it penalizes all concentration, whereas some of that
concentration may be epistemically valuable.
A refinement that separately identifies informed from uninformed large-trader
flow remains an important direction for future work, along the lines proposed
by \citet{rothschild2026heterogeneity}.

\subsection{Platform Design and the Accuracy-Authority Inversion}
\label{subsec:accuracy-authority}

\begin{figure}[H]
  \centering
  \includegraphics[width=0.75\textwidth]{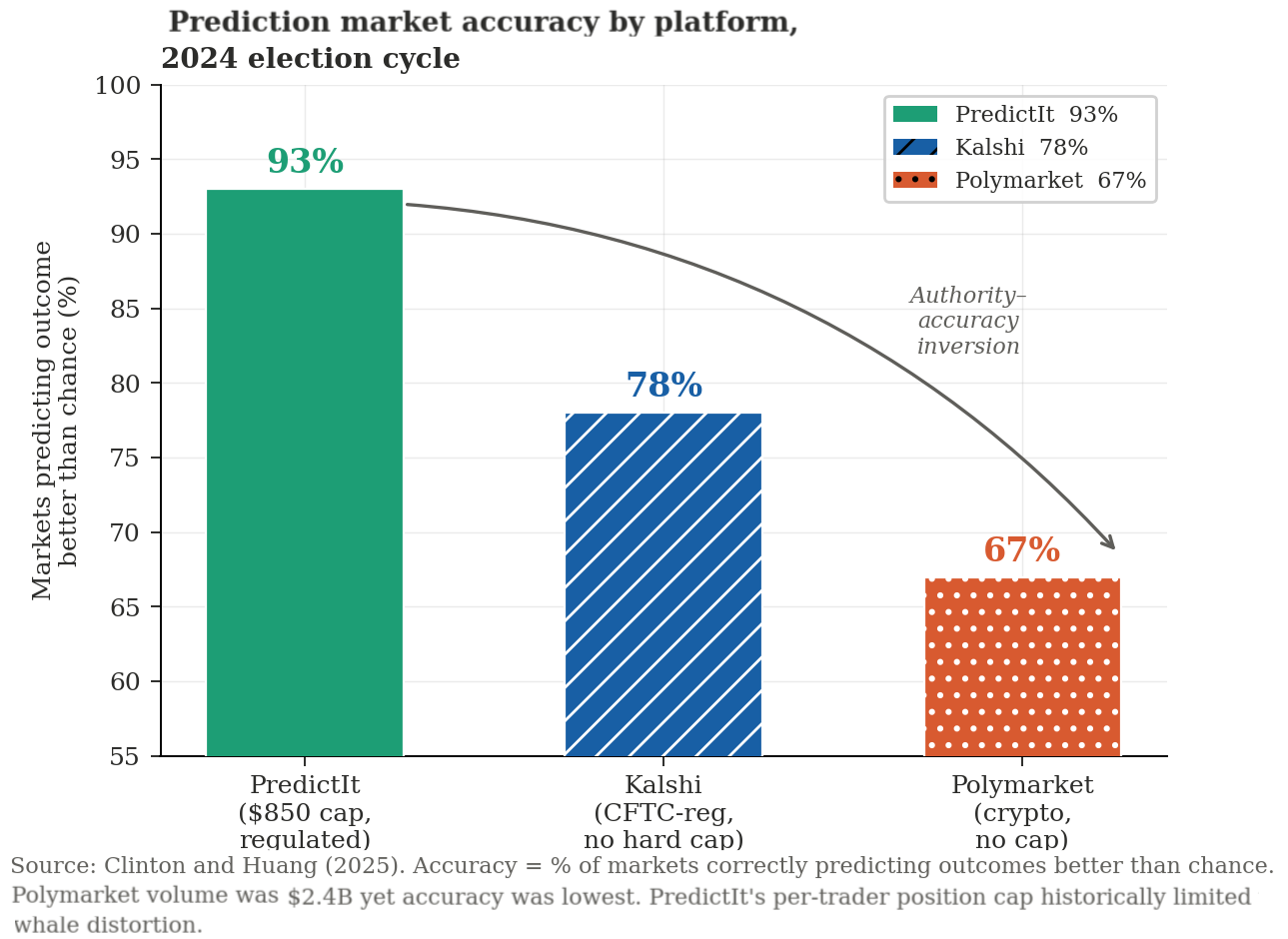}
  \caption{Prediction market accuracy across platforms in the 2024
    election cycle. PredictIt outperforms both Kalshi and Polymarket
    despite far lower volume and tighter position caps.
    Source: \citet{clinton2025accuracy}.}
  \label{fig:accuracy}
\end{figure}

\citet{clinton2025accuracy} analyze over 2,500 prediction markets with
approximately \$2.4 billion in trading volume across the 2024 election cycle.
Their headline finding is summarized in Table~\ref{tab:accuracy} and Figure~\ref{fig:accuracy}.

\begin{table}[H]
  \centering
  \caption{Prediction market accuracy by platform, 2024 election cycle}
  \label{tab:accuracy}
  \setstretch{1.2}
  \small
  \begin{tabular}{lccc>{\raggedright\arraybackslash}p{3.2cm}}
    \toprule
    \textbf{Platform} &
    \textbf{Accuracy (\%)} &
    \textbf{Volume (USD)} &
    \textbf{Position cap} &
    \textbf{Structure} \\
    \midrule
    PredictIt  & 93 & Low        & \$850 (historical) & Regulated, capped \\
    Kalshi     & 78 & Medium     & None (regulated)   & CFTC-registered DCM \\
    Polymarket & 67 & \$2.4B     & None               & Decentralized/crypto \\
    \bottomrule
  \end{tabular}
  \par\smallskip
  \raggedright\footnotesize
  \textit{Notes:} Accuracy measured as the percentage of markets correctly
  predicting outcomes better than chance (directional accuracy).
  Source: \citet{clinton2025accuracy}.
  ``Position cap'' refers to the per-trader limit operative during the
  observation period.
  Polymarket volume figure includes exchange-equivalent and gross activity
  following the decomposition methodology of \citet{tsang2026anatomy}.
\end{table}

The inversion is analytically central.
PredictIt --- the lowest-volume, lowest-visibility, most tightly capped
platform --- produced the most accurate forecasts.
Polymarket --- the highest-volume, most widely cited platform --- produced the
least accurate forecasts.
This does not prove that position caps cause accuracy improvements (the
platforms differ along many dimensions simultaneously), but it is consistent
with the hypothesis that per-trader limits preserve forecasting quality by
preventing concentrated positions from dominating visible prices.

For the paper's argument, the inversion has a direct implication: the platform
most likely to function as a coordination device (high visibility, widely cited,
cross-platform price leader per \citealt{ng2026price}) is simultaneously the
platform least justified in claiming superior predictive content.
Social authority and epistemic robustness are not only decoupled --- they are
inversely ordered across the major platforms.

The \citet{sethi2025prediction} profitability test provides a complementary
perspective.
Rather than measuring raw directional accuracy (which, as they argue, penalizes
confident correct predictions and rewards cautious ones), they ask whether a
trader who believes a forecast and bets accordingly would be profitable.
By this criterion, prediction markets aggregate information more efficiently
than statistical polling models across a broad range of electoral contests.
The two sets of findings are not contradictory: the Clinton-Huang measure
captures \emph{calibration on individual markets}, while the Sethi-Seager
measure captures \emph{relative information advantage over model-based
alternatives}.
Both are relevant.
For coordination purposes, the Clinton-Huang accuracy measure is more important:
a poorly calibrated market can still be socially influential, but it is more
likely to generate self-defeating dynamics when sophisticated actors notice the
miscalibration.

\subsection{Oracle and Resolution Risk}
\label{subsec:oracle}

A fourth source of fragility is specific to decentralized platforms.
In markets that rely on token-based governance to settle disputed outcomes,
the displayed probability reflects not only beliefs about the underlying event
but also expectations about how that event will be adjudicated.
If settlement can be influenced by concentrated voting power, then a market
price encodes both an event probability and an oracle-governance risk premium ---
and these two components are invisible to casual observers who read the
displayed number as a pure probability.

This creates a unique vulnerability in the social-signal framework. A price
can be ``correct'' in the sense of predicting how the oracle will resolve a
contract, while still being systematically biased relative to the true event
probability if the oracle is subject to capture.
The 2025 episodes involving resolution disputes on high-profile geopolitical
markets illustrate that this risk is not theoretical.
For the coordination argument, oracle risk implies that the information
contained in a publicly visible prediction-market price may be contaminated
by an unobservable governance noise component, further widening the gap between
the displayed number and any legitimate probability estimate.

\section{The Self-Fulfilling Loop}
\label{sec:selffulfilling}

The self-fulfilling prophecy is often described too loosely.
In this context, it does not mean that pricing an outcome makes it happen.
It means that a public forecast can alter behavior in ways that increase the
probability of the forecasted outcome.
As Figure~\ref{fig:loop} and Proposition~\ref{prop:threshold} together imply, the key question is not whether a market emits a signal, but whether that signal
achieves $\SCI_s > \tau$ and thereby becomes common knowledge durable enough
to organize behavior.

The mechanism can be stated in four steps:
\begin{enumerate}[noitemsep]
  \item The market emits a public signal that an outcome is becoming more or
        less likely.
  \item Relevant audiences interpret that signal as evidence of informed
        consensus or changing viability, invoking Equation~\eqref{eq:aggregate-response}.
  \item They adjust behavior: traders reposition, journalists reframe, donors
        reallocate, institutions prepare, political actors revise strategy.
  \item Those behavioral changes affect the underlying environment, feeding
        back through Equation~\eqref{eq:market-update}.
\end{enumerate}
When that sequence holds, the forecast is no longer merely describing
expectations; it is participating in the system that produces the outcome.

\paragraph{Finance.}
The strongest evidence comes from finance, where behavior is observable and
response times are short.
FedWatch-style rate expectations are the clearest case.
After a hotter-than-expected inflation print in April 2024, the implied
probability of a June Federal Reserve rate cut fell from roughly 50\% to 19.4\%
in a day; after weaker labor data, the implied probability of a 50-basis-point
cut jumped from 28\% to 49\% in a day \citep{reuters2024fedwatch}.
These shifts did not determine Fed policy.
But traders, firms, and journalists used such probabilities to reprice risk,
hedge exposures, and update policy narratives.
The reflexive channel is strongest here because the audience (financial market
participants) is both the observer of the signal and the primary determinant
of the outcomes the signal is predicting.

\paragraph{Politics.}
The political case is more moderate.
The system is noisier, the causal chain harder to isolate, and the counterfactual
--- what would have happened without the market signal --- harder to identify.
A market move suggesting that a nominee is no longer viable may encourage donors
to pause contributions, operatives to hedge, party elites to discuss alternatives,
and reporters to frame the situation as deteriorating.
Those reactions can then worsen the candidate's position.
The Biden withdrawal market is the clearest recent case.
A jump from 19\% to 50\% within roughly 24 hours is a public reclassification
of the scenario from unlikely to plausible.
But as Table~\ref{tab:shocks} documents, the debate shock that triggered this
jump had a low SCI --- the initial repricing largely reversed --- which means
the Biden case is better described as a case where \emph{social salience}
outran \emph{signal quality}, rather than as an archetypal reflexive loop.

\paragraph{The counterfactual problem.}
A persistent challenge for the reflexivity thesis is the counterfactual.
Donors, media organizations, and political operatives respond to many overlapping
signals simultaneously --- polls, party internal data, media narratives, debate
performance, and prediction markets.
Disentangling the marginal contribution of prediction market prices from this
broader informational environment requires either natural experiments (events
that shift market probabilities exogenously of any change in fundamentals) or
structural identification strategies that are not yet available in the existing
literature.
This limitation is not unique to this paper; it is the central methodological
challenge for any empirical investigation of reflexivity in social systems.
We flag it explicitly because any strong causal claim in this and the two following sections should be read as a conditional claim: ``\emph{if} prediction market signals contribute
to the informational environment, \emph{then} their persistence and credibility
conditions determine whether that contribution is socially consequential.''
The magnitude of the effect remains an open empirical question.

\paragraph{Scope conditions revisited.}
Combining the theoretical framework with the empirical evidence, the scope
conditions for self-fulfilling dynamics can now be stated precisely.
Self-fulfilling effects are most likely when:
\begin{itemize}[noitemsep]
  \item $\SCI_s > \tau$ (signal is credible and durable by microstructure standards);
  \item the Chernov et al.\ event factor loads heavily across multiple
        electoral units, amplifying the coordination value of the signal;
  \item the responding audience is dominated by elites and media rather than
        voters (high $\sigma_{\text{elites}}$, $\sigma_{\text{media}}$);
  \item no dominant rival signal (polls, party surveys, expert consensus)
        provides a competing focal point; and
  \item the platform emitting the signal has high narrative dominance ---
        it is the one that journalists and commentators are most likely to cite.
\end{itemize}
Self-defeating effects are most likely when perceived inevitability suppresses
turnout or mobilization among the favored side, or when the signal's high
visibility galvanizes counter-mobilization among the trailing side.
These two possibilities are examined in Section~\ref{sec:selfdefeating}.
 
\bigskip
 
\section{The Self-Defeating Counterpart}
\label{sec:selfdefeating}
 
Every self-fulfilling mechanism has a theoretically symmetric counterpart.
A strong public probability can reinforce the favored outcome; it can equally
undermine it by changing incentives asymmetrically.
Supporters of the apparent front-runner may become complacent, reduce campaign
effort, or discount the importance of voting.
Supporters of the trailing side may become more alarmed, more mobilized, and
more willing to coordinate against what now looks like an impending loss.
Public probabilities do not push behavior in only one direction; they alter
incentives, and the direction of the response depends on whether the signal
\emph{demobilizes} the apparent winner or \emph{galvanizes} the apparent loser.
 
Brexit and the 2016 U.S.\ presidential election offer suggestive cases.
Both featured highly visible probabilities pointing strongly in one direction ---
roughly 75\% for Remain on the day of the Brexit referendum, and 84--90\% for a
Clinton victory in the final weeks of October 2016 --- yet both ended the other
way.
This does not \emph{prove} that forecast confidence became self-defeating.
The two cases are observational; many factors contributed to both outcomes.
But they establish a pattern that the reflexivity literature needs to take
seriously: the same public signal can generate bandwagon effects, complacency
effects, counter-mobilization, or some mixture of all three, and separating
those channels empirically is often impossible.
 
From the formal framework of Section~\ref{sec:formal}, the self-defeating case
arises when the sign of the behavioral response $B_t$ is negative with respect
to $O_t$ --- that is, when the public signal that outcome $O^*$ is likely
actually reduces the probability of $O^*$ by mobilizing opposing behavior.
The equilibrium condition becomes:
\begin{equation}
  \frac{\partial O_t}{\partial p_t} < 0:
  \quad \text{self-defeating dynamics.}
  \label{eq:selfdefeating}
\end{equation}
 
The sign of this derivative is not observable prior to the fact, and the same
market structure can produce either sign depending on which audience type
dominates the behavioral response.
This is why the empirical record is difficult to interpret, and why
Proposition~\ref{prop:threshold} must be understood as a necessary but not
sufficient condition: a high SCI predicts that the signal will organize
behavior, but it does not specify the direction of that behavioral response.
Identifying the sign requires knowing whether complacency effects or
galvanization effects dominate in the specific context --- a question that
depends on partisan intensity, existing mobilization levels, and the structure
of the media environment.
 
\begin{remark}
  The difficulty of distinguishing self-fulfilling from self-defeating dynamics
  has an important implication for empirical work.
  When a prediction market probability of $p_t = 0.85$ for outcome~$A$ is
  followed by outcome~$B$, it is tempting to attribute the miss to forecast
  error.
  But part of the miss may be \emph{caused} by the forecast: the high
  probability of~$A$ demobilized its supporters while galvanizing~$B$'s.
  This would mean the market was doing its job (reflecting informed consensus)
  while simultaneously contributing to the outcome it was predicting against.
  Disentangling forecast error from reflexive feedback requires a counterfactual
  that does not exist in observational data.
\end{remark}
 
\section{The Power of Perceived Inevitability}
\label{sec:inevitability}
 
Perceived inevitability is the most behaviorally potent form of a public
forecast because it changes the question actors ask.
Instead of asking what outcome they prefer, they begin asking how to position
themselves relative to the outcome they now treat as likely to happen regardless.
This is the point at which a probability stops functioning merely as a forecast
and starts functioning as a coordination cue.
 
The mechanism is cognitive as much as rational.
Many audiences do not process probabilities in a strictly statistical way.
An 80\% probability still implies substantial uncertainty, but in public
discourse it is routinely translated into a narrative of near-certainty.
The step from ``very likely'' to ``basically decided'' is analytically
unjustified but socially common.
Prediction markets intensify this slippage because they package judgment in a
form associated with discipline, competition, and money at risk.
The phrase ``the market gives $X$ an 80\% chance'' sounds less like an opinion
and more like a verdict.
 
The effects differ by audience in ways that parallel the coordination mechanism
described in Section~\ref{sec:social-signals}.
For \textbf{voters}, perceived inevitability can weaken motivation, increase
strategic defection, or reinforce bandwagon impulses.
For \textbf{elites}, the effects are stronger and more operationally
consequential: donors avoid backing what looks like a lost cause, party actors
close ranks around the apparent winner, and institutions begin preparing for
the scenario that now seems dominant.
For \textbf{media systems}, inevitability is especially powerful because it
simplifies narration: a race can be framed as over before it is over, and every
subsequent event is interpreted as confirmation or deviation from the default.
 
Two historical examples illustrate the mechanism.
In the June 2016 Brexit referendum, betting markets implied roughly 75\% for
Remain, and sterling rose as traders positioned for that result.
The visible probability created a shared sense of expected direction and moved
real capital around that expectation, even though the realized outcome ran
against it.
In the 2016 U.S.\ election, major aggregations placed Hillary Clinton's
probability at 84--90\% in mid-October.
Bookmaker Paddy Power reportedly paid out early on a Clinton win --- the
clearest possible illustration of how probability can harden into
institutionalized expectation.
These cases are best understood as demonstrations of the inevitability mechanism
at work in the elite and media channels, regardless of whether they
ultimately produced self-fulfilling or self-defeating outcomes.
 
A probability becomes socially consequential when it no longer merely informs
observers but reorganizes their sense of what counts as realistic, viable, or
worth resisting.
That is the condition under which prediction markets are most likely to shape
the world they are ostensibly forecasting.
And it is also the condition under which the stakes for market design, accuracy,
and regulation are highest --- which is why the governance questions addressed
in Section~\ref{sec:regulation} are substantive rather than procedural.
 
\section{Why This Differs from Direct Manipulation}
\label{sec:manipulation}
 
It is useful to distinguish the coordination mechanism described in this paper
from the model of influence associated with firms such as Cambridge Analytica.
That model emphasized micro-targeting: identify specific individuals, infer
susceptibilities, and deliver customized messages designed to change their
behavior.
Whatever one thinks of its effectiveness, the mechanism was direct,
personalized, and aimed at persuasion one audience segment at a time.
 
The mechanism here is different in kind.
Prediction markets operate at the level of public perception rather than
individualized messaging.
They do not need to persuade each voter, donor, trader, or journalist
separately.
They only need to establish a visible reference point for what the likely path
of events appears to be.
 
That difference matters because macro-level influence can be consequential even
when very few people are directly persuaded.
A shared public signal can alter coordination, timing, and risk assessment across
many actors at once.
Donors can withhold support, journalists can change narrative framing,
institutions can prepare for one scenario rather than another, and traders can
reposition --- not because each has been individually targeted, but because each
is responding to the same publicly legible probability encoded in
Equation~\eqref{eq:ck-signal}.
 
The relevant contrast is therefore not manipulation versus no manipulation, but
\textbf{micro-persuasion} versus \textbf{macro perception-shaping}.
Prediction markets matter, when they matter, by changing the background
assumptions under which collective decisions are made.
That is a quieter mechanism than targeted propaganda, but in coordination-heavy
environments it can be at least as powerful.
 
A fourth source of structural fragility --- oracle and resolution risk --- creates
a further complication here.
If a market's visible probability can be influenced at the resolution layer
(through token-governance capture or disputed adjudication), then a sophisticated
actor can shape public perception without directly trading in large volumes.
They need only influence how an outcome is officially resolved.
This form of influence is entirely invisible to outside observers who see only
the displayed probability, and it does not require the large order flows that the
concentration diagnostics in Section~\ref{sec:microstructure} would flag.
It is therefore both harder to detect and more durable than the whale-trader
narrative that typically dominates regulatory discussions.
 
\section{The AI Amplification Layer}
\label{sec:ai}
 
Artificial intelligence adds a further dimension to the dynamics described in
this paper.
This section should be understood as forward-looking rather than settled; the
empirical literature on AI-mediated prediction-market amplification is still
limited, and the strongest claims here remain speculative.
The grounded point is narrower: AI systems can increase the speed, reach, and
correlation of the behavioral responses that prediction market signals trigger.
 
\paragraph{Signal propagation.}
Automated systems monitor news, ingest market moves, summarize them, and
redistribute interpretations at machine speed.
Concrete examples already exist.
Financial data platforms such as Bloomberg and Refinitiv integrate
market-implied probabilities --- including FedWatch-style indicators and
Kalshi-sourced event contracts --- into dashboards that are increasingly
summarized by AI-assisted analyst tools.
LLM-based portfolio management systems and algorithmic trading agents routinely
query prediction market prices as inputs into allocation decisions.
These systems both consume and react to market signals, creating a feedback loop
in which the same public probability is read, interpreted, and acted upon by
multiple automated agents in near-simultaneous time.
 
\paragraph{Correlated response.}
The second change is the potential for synchronized behavior.
If many trading systems, dashboards, automated summary tools, and recommendation
layers rely on the same public inputs --- and if those inputs include prediction
market prices --- they may generate similar outputs at roughly the same time.
This correlated response does not require any coordination among the systems.
It follows automatically from the shared observation of the same focal signal.
In the formal framework, AI amplification increases the effective $\omega_i$
weights for the fastest-responding audience types (traders, institutional
analysts) while compressing the time between market move and behavioral response.
 
\paragraph{Feedback compression.}
Together, faster propagation and correlated response mean that the feedback
loop in Equation~\eqref{eq:market-update} operates on a shorter time horizon.
If $B_t$ propagates back into $p_{t+1}$ more quickly than human actors alone
would achieve, then the self-fulfilling dynamics described in
Section~\ref{sec:selffulfilling} can close faster, and the window for corrective
information to interrupt the loop narrows.
This is particularly concerning in domains where event-time is short --- geopolitical
crises, central bank announcements, corporate disclosures --- and where a market
move toward one outcome can plausibly shift real-world probabilities before
independent verification is possible.
 
\paragraph{A note on insider trading and privileged information.}
The combination of AI-enabled rapid response and the high stakes of geopolitical
prediction markets creates a risk that has recently received legislative
attention: the possibility that prediction market trades are placed using
privileged government information.
If a visible prediction market probability moves sharply before a public
announcement --- and if that move is then amplified by AI-mediated signal
propagation --- the social consequence is not merely an accuracy failure but a
channel through which classified information is converted into public
coordination signals.
This concern is not theoretical.
\citet{rabinovitz2026markets} --- writing broadly about prediction markets as a threat to democratic information infrastructure --- documents that a Polymarket trade placed shortly
before the U.S.\ military detention of Venezuelan President Maduro in 2026 prompted
legislative attention, as did unusual trading activity preceding airstrikes on
Iran.
In the formal framework, this scenario represents a case where $\nu_t$ in
Equation~\eqref{eq:market-update} is not random noise but \emph{structured
information leakage} --- precisely the situation in which a market signal is
most dangerous as a coordination device, because it is simultaneously most
likely to be accurate and most likely to be acting on information that should
not be publicly available.
 
\section{Real-World Implications}
\label{sec:implications}
 
The practical implication is not that prediction markets should be dismissed.
They can aggregate information usefully and, in appropriate designs and
institutional contexts, outperform conventional forecasting tools
\citep{berg2008prediction,snowberg2013prediction,sethi2025prediction}.
The narrower and more important point is that they should no longer be
understood as mere mirrors.
In domains where expectations shape action, public probabilities can become
part of the environment in which decisions are made.
 
\paragraph{Politics.}
Prediction markets can function as informal coordination devices.
When Polymarket odds of Biden withdrawing moved from approximately 19\% before
the June 27, 2024 debate to around 50\% by the following day, the shift did not
just register sentiment.
It became part of how donors, party actors, and journalists interpreted
viability \citep{wsj2024biden}.
The market did not decide the outcome, but it helped structure the conversation
within which political actors responded.
The SCI analysis shows that this signal's low persistence limited its status as
a durable coordination anchor --- but its vivid social salience still generated
short-term elite repositioning that was consequential.
The Biden withdrawal eventually occurred; whether, and to what degree, the
prediction market signal accelerated that decision is a causal question that
the available evidence cannot answer definitively.
What the evidence can establish is that the signal was present, widely observed,
and embedded in the informational environment at the decisive moment.
 
\paragraph{Finance.}
The reflexive channel is strongest in finance because reaction is faster and
easier to observe.
FedWatch-style probabilities do not merely summarize expectations about the
Federal Reserve; they help organize pricing, hedging, and commentary around
those expectations \citep{reuters2024fedwatch}.
Here the loop is most clearly closed: public odds shape behavior because market
participants explicitly treat them as inputs into their own decisions, and those
decisions collectively determine the market environment the Fed responds to.
 
\paragraph{Geopolitics.}
In geopolitics, the narrative channel matters even when the behavioral feedback
is harder to measure.
Public odds on escalation, ceasefires, leadership changes, or territorial
outcomes influence how governments, firms, media organizations, and outside
observers frame what is becoming plausible.
Which scenarios appear serious enough to prepare for and which are dismissed as
noise is partly determined by the visible market probability --- a mechanism
that operates whether or not the market is accurate.
The insider-trading concerns documented by \citet{rabinovitz2026markets} suggest
that prediction markets are already embedded in the information environment
surrounding high-stakes geopolitical decisions, with consequences that are only
beginning to be understood.
 
\section{Regulation and Institutional Stakes}
\label{sec:regulation}
 
If prediction markets are not merely mirrors but potential participants in
coordination processes, then regulation is not a side issue.
It becomes part of the practical stakes of the argument.
The question is no longer only whether markets forecast well.
It is also whether, where, and under what rules they are allowed to operate
once their probabilities begin to matter for behavior beyond the trading venue.
 
\paragraph{The current landscape.}
That point is no longer abstract.
The Kalshi litigation established federal court precedent for election contracts
under the Commodity Exchange Act; the CFTC under Chair Selig subsequently
withdrew a proposed ban on political and sports-related contracts and signaled
new rulemaking.
State regulators in Nevada, New Jersey, Maryland, and elsewhere have
simultaneously pursued cease-and-desist orders, and a federal-state preemption
dispute is working through multiple circuit courts.
This is an active, unresolved governance contest whose outcome will determine
the institutional architecture within which prediction market signals acquire
(or are denied) social authority.
 
\paragraph{Platform design as regulatory variable.}
The accuracy-authority inversion documented in Section~\ref{sec:microstructure}
has a direct regulatory implication: platform design choices --- position caps,
trader verification, contract categories, resolution mechanisms --- are not
merely market integrity decisions.
They are choices about which signals become socially authoritative.
A licensed and widely quoted market can become more than an information
aggregator; it can become an institutionally recognized reference point for
journalists, donors, firms, and officials who are searching for a credible
summary of uncertain futures.
 
The evidence from \citet{clinton2025accuracy} suggests that PredictIt's
historically tighter position limits preserved both accuracy and the
participatory breadth that grounds democratic legitimacy.
If that finding is robust, it implies that the deregulatory trajectory of
2025--2026 --- removing caps, expanding contract categories, admitting
institutional traders with unlimited position sizes --- may systematically
improve market liquidity while degrading the epistemic quality of the public
signal.
Regulators who care about prediction markets as democratic information
infrastructure should weigh that trade-off explicitly.
 
\paragraph{Insider trading and democratic integrity.}
Beyond the design question, the insider-trading risk identified in
Section~\ref{sec:ai} requires a dedicated regulatory response.
Standard securities law does not apply to most prediction market trades.
State gambling regulation, which could apply, does not have the detection
infrastructure for the specific risk pattern.
The CFTC's derivatives framework does apply to registered platforms, but
the agency's experience with the specific problem of trading on privileged
government information in event-contract markets is limited.
\citet{rabinovitz2026markets} argues, in the most direct treatment of this
problem to date, that prediction markets operating on sensitive political and
military events create financial incentives to leak or exploit classified
information --- incentives that grow as market liquidity deepens.
In the formal framework, this means that $\nu_t$ in
Equation~\eqref{eq:market-update} is systematically contaminated in
high-stakes geopolitical markets, and no amount of platform design improvement
can fully resolve that contamination without access controls at the information
source.
 
\paragraph{Governance as part of the reflexive system.}
Regulation, in other words, does not sit outside the reflexive story told in
this paper.
Oversight decisions help determine which public signals gain legitimacy,
liquidity, and reach.
Choices about market design, disclosure requirements, position limits, and
permissible contract categories shape not only who may trade, but which kinds of
expectations become socially authoritative.
Whether a prediction market remains a forecasting device or begins to function
as a broader coordination signal is not determined solely by the market itself;
it is also determined by the institutional conditions under which that market
operates.
 
\section*{Conclusion}
\addcontentsline{toc}{section}{Conclusion}
 
Prediction markets are not science-fiction machines that bend reality at will.
They are not uniformly reliable, not every price move after a shock is equally
informative, and platform design matters enormously for whether a market is
revealing information or registering temporary pressure.
Some shocks produce persistent repricing; others produce reversal or strongly
two-sided disagreement.
A visible probability should not be read as a transparent verdict on collective
belief simply because it moved quickly \citep{snowberg2013prediction,
tsang2026political}.
 
This paper has argued that the more important question is not whether
prediction markets forecast accurately, but when they become socially
consequential enough to help organize the behavior that determines outcomes.
The Signal Credibility Index --- integrating persistence ($\VR(6)$),
two-sidedness ($\TS$), and concentration ($\HHI$) (see Figures~\ref{fig:paths}--\ref{fig:accuracy}) --- provides a
microstructure-grounded criterion for answering that question.
Applied to the 2024 data, it shows that three major political shocks generated
qualitatively distinct signal types, and that only one (the assassination
attempt) met the conditions for strong coordination effects.
 
The authority paradox is this paper's most striking empirical finding:
the platform with the highest volume and visibility was the least accurate
forecaster, and the market was most socially influential precisely when it was
least epistemically mature.
Social credibility is front-loaded; market quality is back-loaded.
The decoupling of these two properties --- authority and accuracy --- is not a
correctable market failure.
It is a structural feature of how public focal points acquire social force
before they can be tested for reliability.
 
The self-fulfilling and self-defeating possibilities are theoretically symmetric,
and the empirical record cannot cleanly distinguish between them in the cases
most commonly cited.
Brexit and the 2016 U.S.\ election show that high public probabilities can
precede unexpected outcomes.
Whether those probabilities helped produce the unexpected outcome (self-defeating)
or were simply wrong (forecast error) requires counterfactual identification
strategies that the available data do not support.
This limitation should discipline the strong claims in the reflexivity
literature, including the conditional claims in this paper.
 
What prediction markets have done, and what this framework explains, is
transform a category of private expectations into publicly legible common
knowledge that actors use as a reference point for coordination.
That transformation is consequential regardless of whether any particular market
moves behavior in the direction it predicts.
It changes the informational environment --- the background assumptions under
which voters, donors, journalists, traders, officials, and institutions act ---
and therefore the field within which outcomes are decided.
 
Regulation, platform design, and the institutional rules governing which
markets acquire social authority are therefore not peripheral concerns.
They are part of the reflexive system itself.
Once a forecast acquires the status of a credible coordination signal, it
participates in organizing the field of action before the outcome is settled.
How we govern those signals determines, in part, which futures become thinkable
and which are dismissed as noise before they have the chance to materialize.
 
\newpage
\bibliographystyle{plainnat}
\bibliography{refs}
 
\newpage
\appendix
 
\section*{Appendix A: Derivation of the Signal Credibility Index}
\addcontentsline{toc}{section}{Appendix A: Derivation of the Signal Credibility Index}
\setcounter{equation}{0}
\renewcommand{\theequation}{A.\arabic{equation}}
\renewcommand{\theHequation}{A.\arabic{equation}}
 
The Signal Credibility Index introduced in the main text combines three
microstructure diagnostics into a single composite measure.
This appendix provides the derivation and discusses the properties of each
component.
 
\subsection*{A.1 The Variance Ratio Component}
 
The variance ratio $\VR(6)$ tests for deviations from a random walk over a
six-period horizon (30 minutes versus 5 minutes, following the binning
convention in \citealt{tsang2026political}).
Under the null of a random walk, $\VR(6) = 1$.
The ratio is computed as:
\begin{equation}
  \VR(6)_s = \frac{\hat{\sigma}^2_{30}}{6\,\hat{\sigma}^2_5},
  \label{eq:vr-appendix}
\end{equation}
where $\hat{\sigma}^2_k$ denotes the sample variance of $k$-minute log-price
changes in the post-shock window $[t_s, t_s + 4\text{h}]$.
 
For the coordination argument, the relevant property is:
\begin{itemize}[noitemsep]
  \item $\VR(6)_s > 1$: momentum --- each 5-minute move predicts further movement
    in the same direction, consistent with informed trading and durable updating.
    High coordination potential.
  \item $\VR(6)_s < 1$: reversal --- each 5-minute move is partially corrected
    in subsequent periods, consistent with temporary liquidity pressure.
    Low coordination potential.
  \item $\VR(6)_s \approx 1$: unpredictable short-horizon dynamics.
    Ambiguous coordination potential.
\end{itemize}
 
\subsection*{A.2 The Two-Sidedness Component}
 
The two-sidedness index $\TS_t$ (Equation (7) in the main text) measures the
degree to which post-shock trading is directionally one-sided.
For coordination purposes, low two-sidedness (consensus) is the
coordination-supporting state, while high two-sidedness (disagreement)
is the coordination-suppressing state.
The component therefore enters the SCI as $(1 - \TS_s)$, ranging from 0
(maximum disagreement) to 1 (maximum consensus).
 
\subsection*{A.3 The Concentration Component}
 
The Herfindahl-Hirschman Index of trader position sizes is:
\begin{equation}
  \HHI_s = \sum_{j=1}^{N_s} \left(\frac{v_j}{\sum_{j'} v_{j'}}\right)^2,
  \label{eq:hhi-appendix}
\end{equation}
where $v_j$ is the notional position size of trader $j$ in the post-shock window
and $N_s$ is the number of active traders.
$\HHI_s \in [1/N_s, 1]$, with higher values indicating more concentrated
positions.
The component $(1 - \HHI_s)$ ranges from 0 (maximum concentration) to
$(1 - 1/N_s) \approx 1$ for large $N_s$.
 
\subsection*{A.4 Properties of the Composite SCI}
 
The composite SCI is the product of the three components:
\begin{equation}
  \SCI_s = \VR(6)_s \times (1 - \TS_s) \times (1 - \HHI_s).
  \label{eq:sci-appendix}
\end{equation}
 
By construction, $\SCI_s \geq 0$, with higher values indicating greater signal
credibility.
The multiplicative structure implies that a deficiency in any one component
substantially reduces the composite: a highly persistent signal from a
concentrated trader pool (low $1 - \HHI$) scores low despite its persistence.
A consensus signal (low $\TS$) that reverses quickly (low $\VR(6)$) also scores
low.
 
\begin{remark}
  A limitation of the SCI as specified is that it treats all three components
  as equally weighted.
  In principle, domain-specific evidence could justify differential weighting
  --- for example, giving greater weight to persistence in financial markets
  (where the reflexive loop closes faster) versus political markets (where the
  relevant behavioral responses may be slower).
  Estimating optimal weights is an empirical question that requires panel data
  on behavioral responses across multiple shocks, which is not yet available.
\end{remark}
 
\subsection*{A.5 Calibration Against the 2024 Data}
 
The approximate SCI values in Table 1 of the main text are derived from the
qualitative descriptions of $\VR(6)$ and two-sidedness in
\citet{tsang2026political}, combined with the concentration evidence from
\citet{tsang2026anatomy} (which documents declining $\lambda$ over the
election cycle as a proxy for declining effective $\HHI$).
Because $\HHI_s$ is not directly reported in either paper, the concentration
component is held constant across the three events (treated as moderate-high for
all events in the June-July window, consistent with the pre-maturation phase
of the market described in \citealt{tsang2026anatomy}).
 
Future work should estimate $\SCI_s$ directly from blockchain transaction logs,
which provide the necessary position-level data.
 
\newpage
\section*{Appendix B: Summary of Key Variables and Notation}
\addcontentsline{toc}{section}{Appendix B: Summary of Key Variables and Notation}
 
\begin{table}[H]
  \centering
  \caption*{Table B.1: Summary of notation used in the formal framework}
  \label{tab:notation}
  \setstretch{1.2}
  \small
  \begin{tabular}{lll}
    \toprule
    \textbf{Symbol} & \textbf{Definition} & \textbf{Introduced} \\
    \midrule
    $p_t$             & Market-implied probability at time $t$              & Sec.~2 \\
    $O_t$             & Underlying outcome variable at time $t$             & Sec.~2 \\
    $\mathbf{X}_t$    & Vector of exogenous fundamental drivers             & Eq.~(1) \\
    $\varepsilon_t$   & Fundamental noise term                              & Eq.~(1) \\
    $\sigma(p_t)$     & Volatility of market signal over relevant horizon   & Eq.~(3) \\
    $\delta$          & Minimum behavioral threshold for social consequence & Eq.~(3) \\
    $\beta_i$         & Behavioral response function of audience $i$        & Eq.~(4) \\
    $\sigma_i$        & Audience $i$'s sensitivity to market signals         & Eq.~(4) \\
    $\omega_i$        & Outcome-generating weight of audience $i$           & Eq.~(4) \\
    $B_t$             & Aggregate behavioral response at time $t$           & Eq.~(4) \\
    $\Delta I_t$      & Exogenous information shock at time $t$              & Eq.~(5) \\
    $\nu_t$           & Market microstructure noise term                    & Eq.~(5) \\
    $p_t^{(k)}$       & Market price on platform $k$ at time $t$            & Eq.~(10) \\
    $w_k(t)$          & Audience-weighted citation share of platform $k$    & Eq.~(10) \\
    $p_t^{*}$         & Effective common-knowledge signal at time $t$       & Eq.~(10) \\
    $\VR(6)_s$        & Variance ratio for shock $s$ (30-min/5-min)        & Eq.~(6) \\
    $\TS_s$           & Two-sidedness index for shock $s$                   & Eq.~(7) \\
    $\HHI_s$          & Herfindahl-Hirschman Index for shock $s$            & App.~A \\
    $\SCI_s$          & Signal Credibility Index for shock $s$              & Eq.~(9) \\
    $\tau$            & SCI threshold for coordination effects              & Prop.~1 \\
    \bottomrule
  \end{tabular}
\end{table}

\end{document}